\documentclass[showpacs,showkeys,superscriptaddress]{revtex4-2}
\usepackage{amsmath}
\usepackage{amssymb}
\usepackage{graphicx}
\usepackage[hypertex]{hyperref}

\begin{document}
\title{Static and collapsing configurations supported by the spinor fluid
 }

\author{
Vladimir Dzhunushaliev
}
\email{v.dzhunushaliev@gmail.com}
\affiliation{
Department of Theoretical and Nuclear Physics,  Al-Farabi Kazakh National University, Almaty 050040, Kazakhstan
}

\affiliation{
Institute of Nuclear Physics, Almaty 050032, Kazakhstan
}


\affiliation{
Academician J.~Jeenbaev Institute of Physics of the NAS of the Kyrgyz Republic, 265 a, Chui Street, Bishkek 720071, Kyrgyzstan
}

\author{Vladimir Folomeev}
\email{vfolomeev@mail.ru}
\affiliation{
Institute of Nuclear Physics, Almaty 050032, Kazakhstan
}
\affiliation{
Academician J.~Jeenbaev Institute of Physics of the NAS of the Kyrgyz Republic, 265 a, Chui Street, Bishkek 720071, Kyrgyzstan
}
\affiliation{
International Laboratory for Theoretical Cosmology, Tomsk State University of Control Systems and Radioelectronics (TUSUR),
Tomsk 634050, Russia
}

\begin{abstract}
We study a gravitating spherically symmetric nonrelativistic configuration consisting of a spinor fluid
whose effective equation of state is derived from a consideration of a limiting system supported by a massive nonlinear spinor field.
For such a configuration, we find a two-parametric family of static regular solutions describing compact objects
whose masses and sizes are determined by the central density of the spinor fluid and the mass of the spinor field.
Using the similarity method, we consider a gravitational collapse of an initially uniform system consisting of the spinor fluid.
We estimate the general characteristics of the collapse and show that a resulting nonuniform distribution of the fluid occurs
that may serve as a core for the creation of equilibrium starlike configurations.
\end{abstract}

\maketitle

\section{Introduction}
Compact gravitating configurations supported by fundamental fields with different spins have
been the object of vigorous investigations in recent years. The main emphasis in the literature is being placed  on studying
the so-called boson stars~--  objects supported by scalar (spin-0) fields. The characteristics of such fields may
lie within a very wide range, ensuring a wide variety of physical properties of the resulting objects. In particular, their total masses may vary within a range
from those which are typical of atoms up to the masses comparable to those of galaxies~\cite{Schunck:2003kk,Liebling:2012fv}.

Gravitating configurations consisting of fields with nonzero spins have been considerably less studied. It is worth pointing out here the objects supported by
massive vector (spin-1) fields (Proca stars~\cite{Brito:2015pxa,Herdeiro:2017fhv,Sanchis-Gual:2019ljs}) and massless vector fields (Yang-Mills systems~\cite{Bartnik:1988am,Volkov:1998cc}).
In the case of fractional-spin fields it is also possible to obtain spherically symmetric systems. In the presence of strong gravitational fields, the literature in the field offers
both the systems consisting of linear
 \cite{Finster:1998ws,Finster:1998ux,Herdeiro:2017fhv} and nonlinear spin-1/2 fields~\cite{Krechet:2014nda,Adanhounme:2012cm,Bronnikov:2019nqa}.
Nonlinear spinor fields are also used to study cylindrically symmetric solutions~\cite{Bronnikov:2004uu}, wormhole solutions~\cite{Bronnikov:2009na},
and various cosmological problems (see Refs.~\cite{Ribas:2010zj,Ribas:2016ulz,Saha:2016cbu} and references inside).

In our previous works, we have considered gravitating configurations consisting of two nonlinear spin-1/2 fields. In doing so,
the cases both of purely Einstein-Dirac systems~\cite{Dzhunushaliev:2018jhj} and of configurations containing also U(1) Maxwell and Proca fields~\cite{Dzhunushaliev:2019kiy},
 or SU(2) Yang-Mills and Proca fields~\cite{Dzhunushaliev:2019uft} have been examined. The remarkable feature of all such configurations is that,
 for some values of the spinor field coupling constant $\lambda$,
it is possible to obtain configurations with masses comparable to the Chandrasekhar mass and with effective radii of the order of
kilometers. Such objects can already be regarded as self-gravitating Dirac stars which are prevented from collapsing
under their own gravitational fields due to the Heisenberg uncertainty principle.

If the aforementioned  Dirac stars do exist in nature, it is natural to ask how they are formed. When considering ordinary stars consisting of usual matter
(a macroscopic fluid), the question of their formation is reduced to a consideration of a gravitational collapse of some initially rarefied medium having low pressure and temperature.
 In the process of collapse, these quantities grow, and this eventually leads to the formation of a dense and hot system
that can already exist in an equilibrium state for a sufficiently long time (star formation)~\cite{Prialnik}.
For systems supported by fundamental fields, the process of their dynamic evolution has been considerably less studied. In particular,
it is worth noting here the investigations of gravitational collapse of massless scalar~\cite{Choptuik:1992jv,Gonzalez:2008xk} and spinor~\cite{Ventrella:2003fu} fields,
as well as of mixed systems consisting of a massless scalar field nonminimally coupled to a perfect isothermal fluid~\cite{Folomeev:2011uj}.
In general, however, one might expect that a consideration of collapse of fundamental fields is much more complicated than dealing with a hydrodynamic fluid.

In the present paper, we will use the fact that in the case of the Dirac stars under consideration, for physically reasonable values of the coupling constant~$\lambda$, it is possible to
replace the spinor field by a hydrodynamic fluid described by some effective equation of state (EOS). As was demonstrated in Ref.~\cite{Dzhunushaliev:2018jhj},
a description of the Dirac stars using such a fluid agrees well with
a purely field description of such systems. This will permit us to consider the dynamics of such a fluid (the collapse process), instead of dealing with a much more complicated problem
of collapse of a nonlinear spinor field. Moreover, since in this paper we will work in Newtonian gravity, it will be demonstrated that in the limit of weak
gravitational fields the effective EOS can be reduced to a polytropic EOS. The latter possibility will permit us to consider a self-similar motion of the fluid, which greatly simplifies our task.

The paper is organized as follows. In Sec.~\ref{spinor_fluid}, we present the effective equation of state of the spinor fluid under consideration.
This EOS is employed in Sec.~\ref{stat_conf} to construct regular static solutions describing equilibrium starlike configurations with finite masses and sizes.
Using the fact that the EOS can be reduced to a polytropic form, in Sec.~\ref{collapse} we consider a self-similar motion (collapse) of such a fluid and demonstrate how
a nonuniform distribution of the spinor fluid is formed from an initially uniform cloud.
Finally, in Sec.~\ref{conclus}, we summarize the results obtained.


\section{Spinor fluid}
\label{spinor_fluid}

 In Refs.~\cite{Dzhunushaliev:2018jhj,Dzhunushaliev:2019kiy,Dzhunushaliev:2019uft} we have studied gravitating systems with a nonlinear spin-1/2 field $\psi$
 described by the Lagrangian
$$
	L_{\text{sp}} =	\frac{i \hbar c}{2} \left(
			\bar \psi \gamma^\mu \psi_{; \mu} -
			\bar \psi_{; \mu} \gamma^\mu \psi
		\right) - \mu c^2 \bar \psi \psi + \frac{\lambda}{2} \left(\bar\psi\psi\right)^2,
$$
where $\lambda$  is the coupling constant and $\mu$ is the mass of the spinor field.

It was demonstrated in Ref.~\cite{Dzhunushaliev:2018jhj} that if we introduce the dimensionless coupling constant
 $\bar \lambda \sim \left(M_{\text{Pl}}/\mu\right)^2\lambda$, where $M_{\text{Pl}}$ is the Planck mass,
 in the limit of $|\bar \lambda|\gg 1$, it is possible to get configurations whose sizes and masses are comparable to those typical
 of neutron stars.
 It was also shown there that such limiting configurations could be described by some effective EOS relating the pressure
$p$  and the energy density $\varepsilon$ as
\begin{equation}
\label{EoS_eff}
	p = \frac{\varepsilon_0}{9} \left(
		1 + 3\frac{\varepsilon}{\varepsilon_0}-\sqrt{1 + 6\frac{\varepsilon}{\varepsilon_0}}
	\right),
\end{equation}
where $\varepsilon_0=\mu c^2/\lambda_c^3\equiv\mu^4 c^5/\hbar^3$ can be regarded as some characteristic
energy density of the configuration [here $\lambda_c=\hbar/(\mu c)$ is the constant, which
need not be associated with the Compton length since we consider a classical theory].
A configuration described by such an effective EOS may then be regarded as consisting of a fluid, which we will refer to as {\it a spinor fluid}.

Before proceeding to studying the system described by the effective EOS~\eqref{EoS_eff}, it is worth emphasizing the following point.
 In the general case of spherically symmetric configurations supported by spinor fields, there are two different effective pressures~--
 the radial, $p_r$, and tangential, $p_t$. Correspondingly, the system is anisotropic~\cite{Dzhunushaliev:2018jhj}.
 However, using the results of Ref.~\cite{Dzhunushaliev:2018jhj}, it can be shown that in the limit of $|\bar \lambda|\gg 1$
 these two pressures become equal, i.e., the system becomes isotropic. In this connection, in the present paper, we consider only one
 effective pressure $p=p_r=p_t$ appearing in Eq.~\eqref{EoS_eff}.

\section{Static configurations}
\label{stat_conf}

In this section we consider a compact static gravitating configuration consisting of the spinor fluid modeled by the EOS~\eqref{EoS_eff}.
Since we will work in Newtonian gravity throughout the paper, this will permit us to simplify drastically the problem under consideration.
In this case, the corresponding static equations can be written in the form:
\begin{eqnarray}
\label{mass_eq_d}
&&\frac{dM}{dr}	=\frac{4\pi}{c^2}r^2 \varepsilon,\\
\label{energ_eq_d}
&&\frac{dp}{dr}=-\frac{G M(r)}{r^2}\frac{\varepsilon}{c^2},\\
\label{pois_eq_d}
&&\frac{d\varphi}{dr}=\frac{G M(r)}{r^2}.
\end{eqnarray}
Here, $G$ is the Newtonian gravitational constant, $M(r)$ is the current mass enclosed
in a sphere of radius $r$, and $\varphi$ is the Newtonian gravitational potential.

For numerical integration of the above equations, it is convenient to rewrite them in terms of the following dimensionless variables:
\begin{equation}
\label{dmls_var}
x=\sqrt{4\pi}	\frac{r}{R_L}, \quad \bar M=\sqrt{4\pi}	\frac{M}{M_L}, \quad \bar\varepsilon=\frac{\varepsilon}{\varepsilon_0}, \quad \bar\varphi=\frac{\varphi}{c^2}
\quad \text{with} \quad M_L=\frac{M_{\text{Pl}}^3}{\mu^2}, \quad R_L=\lambda_c\frac{M_{\text{Pl}}}{\mu}.
\end{equation}
The parameters $M_L$ and
$R_L$ are the characteristic mass and radius obtained by
Landau in considering compact configurations consisting
of an ultrarelativistic degenerate Fermi gas within the framework of Newtonian gravity.
As a result, taking into account the EOS~\eqref{EoS_eff}, we have from Eqs.~\eqref{mass_eq_d}-\eqref{pois_eq_d} the dimensionless equations
\begin{eqnarray}
\label{mass_eq}
&&\frac{d\bar M}{dx}	=x^2 \bar\varepsilon,\\
\label{energ_eq}
&&\frac{1-\sqrt{1+6\bar\varepsilon}}{3\sqrt{1+6\bar\varepsilon}}\frac{d\bar\varepsilon}{dx}=\frac{\bar M}{x^2}\bar\varepsilon,\\
\label{pois_eq}
&&\frac{d\bar\varphi}{dx}=\frac{\bar M}{x^2}.
\end{eqnarray}

We solve this set of equations numerically subject to the following boundary
conditions given in the vicinity of the center of the configuration $x=0$:
$$
\bar\varepsilon\approx \bar\varepsilon_c+\frac{1}{2}\bar\varepsilon_2 x^2, \quad
\bar M\approx \frac{1}{3}\bar M_3 x^3, \quad \bar\varphi\approx \bar\varphi_c+\frac{1}{2}\bar\varphi_2 x^2,
$$
where $\bar\varphi_c$ and $\bar\varepsilon_c$ denote the central values of the corresponding quantities.
The values of the coefficients $\bar\varepsilon_2$, $\bar M_3$, and $\bar\varphi_2$
can be found from Eqs.~\eqref{mass_eq}-\eqref{pois_eq}.

We start the numerical procedure near the origin $x\approx 0$
and proceed to the boundary point $x=x_b$ where the energy density goes to zero.
The mass contained inside the sphere of the radius $x_b$ will be treated as the total mass of the configuration.
The magnitude of the free parameter $\bar\varphi_c$ is chosen so as to obtain an asymptotically flat spacetime with
 $\bar\varphi \to 0$ as $x\to \infty$. The value of the free parameter $\bar\varepsilon_c$ is arbitrary but it must be chosen so that
 the Newtonian approximation, according to which one must have $|\bar\varphi| \ll 1$ over all space,
 will remain valid.

\begin{figure}[t]
				\includegraphics[width=1\linewidth]{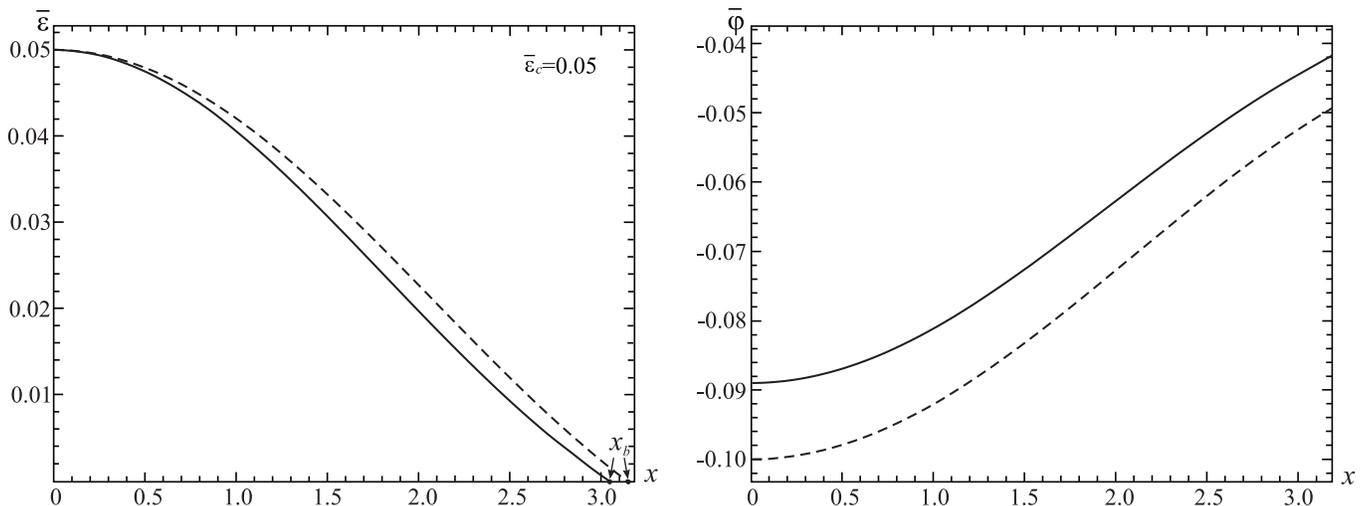}
		\caption{The distributions of the dimensionless energy density of the spinor fluid (left panel) and Newtonian gravitational potential (right panel) for $\bar \varepsilon_c=0.05$.
The solid curves correspond to the system described by the effective EOS~\eqref{EoS_eff} and the dashed curves~-- by the approximate EOS $p=\varepsilon^2/2$.
For larger values of $\bar \varepsilon_c$, the Newtonian approximation ($|\bar\varphi| \ll 1$ over all space) is not already valid.
		}
		\label{fig_stat_conf}
\end{figure}

The numerical calculations indicate that the latter condition is satisfied only in the case when $\bar\varepsilon\lesssim 0.1$ (see Fig.~\ref{fig_stat_conf}).
In this case, the EOS~\eqref{EoS_eff} reduces to the approximate expression
$p\approx\varepsilon^2/2$, for which Eqs.~\eqref{mass_eq_d}-\eqref{pois_eq_d} have the well-known particular solution~\cite{Zeld}
$$
\bar\varepsilon=\bar\varepsilon_c\frac{\sin x}{x}, \quad \bar M=\bar\varepsilon_c\left(\sin x-x\cos x\right), \quad
\bar \varphi=-\bar\varepsilon_c\left(1+\frac{\sin x}{x}\right),
$$
already written in terms of the dimensionless variables~\eqref{dmls_var}. These solutions (shown by the dashed curves in Fig.~\ref{fig_stat_conf}) describe a configuration
whose boundary is located at $x_b=\pi$, thereby defining  the radius, $R$, and the total mass of such a system, $M_{\text{tot}}$, according to~\eqref{dmls_var}, as
\begin{equation}
\label{R_M_approx}
R=\frac{\sqrt{\pi}}{2} R_L, \quad M_{\text{tot}}= \frac{\sqrt{\pi}}{2} \bar\varepsilon_c M_L\quad \text{with} \quad
M_L=1.63 M_\odot\left(\frac{\text{GeV}}{\mu}\right)^2, \quad R_L=2.41\left(\frac{\text{GeV}}{\mu}\right)^2~\text{km},
\end{equation}
where $\bar\varepsilon_c\lesssim 0.1$.  Thus we have a two-parametric family of solutions describing static configurations whose characteristics are entirely determined
by the mass of the spinor field, $\mu$, and by the central energy density of the fluid,~$\bar\varepsilon_c$.

It is seen from the expressions~\eqref{R_M_approx} that, for a fixed value of $\mu$,
there is some maximum total mass.
In particular, for  $\mu \sim 1~\text{GeV}$, we have the mass of the order of the Chandrasekhar mass.
In turn, in the two extreme cases, we have: (i)~for superlight particles of mass $\mu \sim 1~\text{eV}$ (which are used, for example, in modeling
fermionic dark matter~\cite{Narain:2006kx}),
we get very heavy and large configurations of mass
 $M_{\text{tot}} \lesssim 10^{17} M_\odot$ with the radius of the order of a typical
galaxy size, $R\sim 10^{23}~\text{cm}$; and (ii)~for superheavy fermions of mass $\mu \sim 100~\text{GeV}$,
we get light and small objects of mass $M_{\text{tot}} \lesssim 10^{-3} M_\odot$
and  $R\sim 10~\text{cm}$.
The characteristics of objects of this type, if they exist at all, should evidently lie somewhere between these limiting values.

\section{Self-similar collapse}
\label{collapse}

In this section we consider a gravitational collapse of the spinor fluid. The collapse process
has been studied frequently in the literature both in relativistic and nonrelativistic cases. In doing so, two main approaches are used:
(i)~the set of hydrodynamic partial differential equations is solved; and (ii)~a self-similar motion of matter is considered.

Here, we will study self-similar motions of the spinor fluid.
For the EOS~\eqref{EoS_eff}, it is impossible to obtain self-similar solutions.
But since the resulting static configurations can be approximately described by a power-law EOS (see the previous section),
we will employ below the following approximate form of the EOS~\eqref{EoS_eff}:
\begin{equation}
\label{EoS_approx}
p\approx K\rho^\gamma,
\end{equation}
where $\rho\equiv \varepsilon/c^2$ is the mass density of the spinor fluid, and $K$ and $\gamma$ are some constants
whose magnitudes depend on the value of $\rho$.
It follows from Eq.~\eqref{EoS_eff} that there are two limiting cases. First, in the
nonrelativistic case, where $\bar\varepsilon\ll 1$, we get the quadratic dependence, $p\approx \varepsilon^2/2$. Such an EOS has been employed in Sec.~\ref{stat_conf}
in constructing the approximate static solutions.
Second, in the ultrarelativistic case, $\bar\varepsilon\gg 1$, we
have $p\approx \varepsilon/3$.

Self-similar solutions with the EOS~\eqref{EoS_approx} are well known in the literature. In Newtonian gravity, they have been studied in detail, in particular in Ref.~\cite{Suto:1988}.
In doing so, one can use the hydrodynamic equations written in the form:
\begin{eqnarray}
\label{eq_cont_2}
\frac{\partial M}{\partial t}+4\pi r^2 \rho v&=&0,\\
\label{mass_fun_2}
\frac{\partial M}{\partial r}&=&4\pi r^2 \rho,\\
\label{eq_eul_3}
\frac{\partial v}{\partial t}+v \frac{\partial v}{\partial r}&=&-\frac{1}{\rho}\frac{\partial p}{\partial r}-
\frac{G M}{r^2},
\end{eqnarray}
where $M=M(r,t)$ is the total mass inside radius $r$ at time $t$ and $v$ is the fluid velocity.
For the EOS~\eqref{EoS_approx}, the above equations permit the introduction of similarity variables given by~\cite{Suto:1988}:
\begin{align}
\label{self_variab}
\begin{split}
x&=-\frac{r}{\sqrt{k}t},\quad v(r,t)=-\sqrt{k}u(x), \quad \rho(r,t)=\frac{\alpha(x)}{4\pi G t^2}, \\
p(r,t)&=\frac{k}{4\pi G t^2}[\alpha(x)]^\gamma, \quad M(r,t)=-\frac{k^{3/2} }{G}t\,m(x),
\end{split}
\end{align}
where $k$ is some dimensional constant. The above choice of the variables 
implies that the collapse starts at some instant $t<0$ and proceeds with decreasing $|t|$. Also,
one can find from Eq.~\eqref{EoS_approx} that
\begin{equation}
\label{polyt_coeff}
K=k(4\pi G t^2)^{\gamma-1},
\end{equation}
and correspondingly for $\gamma>1$ the parameter $K$ decreases with decreasing $|t|$.

Next, rewriting Eqs.~\eqref{eq_cont_2} and \eqref{mass_fun_2} in terms of the similarity variables~\eqref{self_variab}, one can get an algebraic expression for $m(x)$
in terms of $\alpha(x)$ and $u(x)$,
$$
m(x)= x^2 \alpha(x-u).
$$
Using this expression, Eqs.~\eqref{eq_cont_2}-\eqref{eq_eul_3} can be written out in the final similarity form as
\begin{eqnarray}
\label{eq_alpha}
&&\frac{d\alpha}{dx}=
\frac{x-u}{x}\frac{\left[2u+x(\alpha-2)\right]\alpha}{(x-u)^2-\gamma\alpha^{\gamma-1}},\\
\label{eq_u}
&&\frac{du}{dx}=\frac{x-u}{x}\frac{x\alpha(x-u)-2\gamma\alpha^{\gamma-1}}{(x-u)^2-\gamma\alpha^{\gamma-1}}.
\end{eqnarray}

In constructing collapsing solutions of Eqs.~\eqref{eq_alpha} and \eqref{eq_u}, as initial configurations, the literature in the field offers two types of systems:
(i)~static self-gravitating configurations with a nonuniform distribution of a fluid (see, e.g., Refs.~\cite{Shu:1977uc,Hunter:1977}) and
(ii)~an initially uniformly distributed fluid  (see, e.g., Refs.~\cite{Larson:1969mx,Penston:1969yy}).  Depending on the value of the parameter  $\gamma$, there are qualitatively different types of solutions~\cite{Suto:1988}.

As initial configurations, in the present paper we will use uniform systems. Because of the Jeans instability, the initial configurations whose radii are close to the Jeans length are able to collapse.
For the polytropic EOS~\eqref{EoS_approx}, the Jeans length is
\begin{equation}
\label{Jeans_len}
\lambda_{\text{J}}=\sqrt{\frac{\pi K \gamma}{G}\rho_b^{\gamma-2}},
\end{equation}
where $\rho_b$ is the uniform background density of the fluid. The characteristic size of such an initial system can be taken as $R_i\approx \lambda_{\text{J}}/2$.

As boundary conditions for Eqs.~\eqref{eq_alpha} and \eqref{eq_u}, we choose the following expressions which can be obtained from the expansion of the functions
 $\alpha(x)$ and $u(x)$ in a Taylor series in the vicinity of the center of the configuration $x=0$:
\begin{equation}
\label{bound_cond}
u(x)\approx \frac{2}{3}x-\frac{\alpha_c^{1-\gamma}}{45\gamma}\left(\alpha_c-\frac{2}{3}\right)x^3+\cdots, \quad
\alpha(x)\approx \alpha_c-\frac{\alpha_c^{2-\gamma}}{6\gamma}\left(\alpha_c-\frac{2}{3}\right)x^2+\cdots.
\end{equation}
These expressions are given in terms of the central value $\alpha_c$.

Using these boundary conditions, we integrate numerically Eqs.~\eqref{eq_alpha} and \eqref{eq_u}.
However, this numerical study needs some caution because the solution can pass through critical points~\cite{Suto:1988}.
Their existence implies that the denominator of Eqs. \eqref{eq_alpha} and \eqref{eq_u} vanishes:
\begin{equation}
\label{denom}
(x-u)^2-\gamma\alpha^{\gamma-1}=0.
\end{equation}
Then a necessary condition for the existence of a nonsingular solution is that the numerator in Eq.~\eqref{eq_u} also vanishes:
\begin{equation}
\label{numer}
x\alpha(x-u)-2\gamma\alpha^{\gamma-1}=0.
\end{equation}
This also implies that the numerator in~\eqref{eq_alpha} is simultaneously equal to zero, which can easily be  checked.
Thus Eqs.~\eqref{denom} and \eqref{numer}
are necessary conditions ensuring the existence of nonsingular solutions at a critical point, $x=x_*$. Its location and the corresponding value of the velocity  $u_*$
expressed in terms of $\alpha_*$ are then
$$
x_*=2\sqrt{\gamma}\alpha_*^{(\gamma-3)/2}, \quad u_*=\sqrt{\gamma}\alpha_*^{(\gamma-3)/2}(2-\alpha_*).
$$

There are two possible types of the behavior of solutions: (i)~a solution passes through the critical point, and it is regular and
(ii)~the denominator in Eqs.~\eqref{eq_alpha} and \eqref{eq_u} vanishes at some point but the numerator remains nonzero and finite. This corresponds to singular solutions.
Whether or not a solution passes through the critical point is determined by two free parameters~-- the polytropic exponent $\gamma$
and the central density $\alpha_c$. A solution for the case of $\gamma=2$ in which we are interested here is given in the next subsection.

\subsection{Numerical calculations 
}
\label{num_collap}

In this subsection we give an example of self-similar solutions for the polytropic exponent $\gamma=2$.
In this case, regular solutions to Eqs.~\eqref{eq_alpha} and \eqref{eq_u} with the boundary conditions~\eqref{bound_cond}
do exist only when the solutions pass through the critical point (see above). Aside from this, solutions with decreasing density are present only when the central value
of the density $2/3\leqslant \alpha_c\lesssim 0.927$. The value $\alpha_c=2/3$  corresponds to a strictly uniform distribution of matter:
$\alpha(x)=2/3$ with $u(x)=2/3\, x$ (the homogeneous expansion solution in Newtonian cosmology~\cite{Suto:1988}).
When $\alpha_c<2/3$, the density grows with increasing $x$, which seems physically unacceptable.

Before numerically solving Eqs.~\eqref{eq_alpha} and \eqref{eq_u}, let us write down their asymptotic solution for $x\gg 1$. This is the limit
approached at any fixed value of $r$ as the collapse progresses and $t$ goes to zero.
For such asymptotic limit, we have from Eqs.~\eqref{eq_alpha} and \eqref{eq_u}:
\begin{equation}
\label{asymp_sol}
\alpha\approx \frac{\alpha_\infty}{x^2}, \quad u\approx u_\infty-\frac{2\alpha_\infty}{x^3},
\end{equation}
where $\alpha_\infty$ and $u_\infty$ are integration constants.

\begin{figure}[t]
				\includegraphics[width=.4\linewidth]{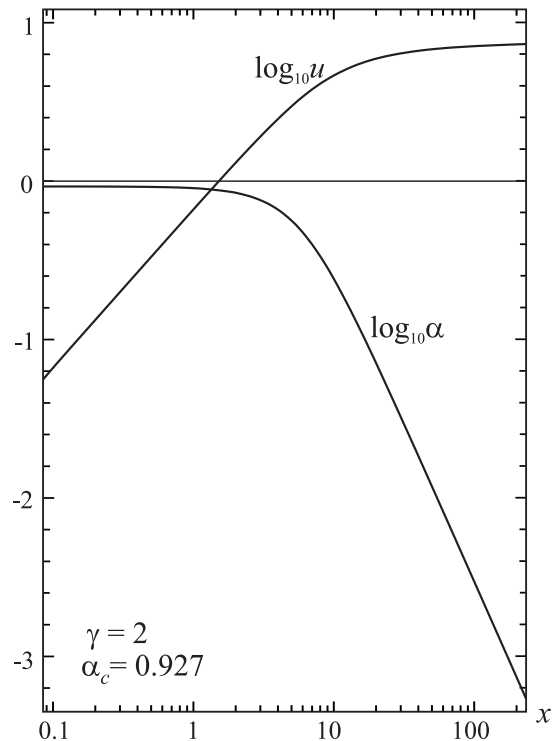}
		\caption{
		Similarity solutions for the velocity $u(x)$ and the density $\alpha(x)$ [see Eq.~\eqref{self_variab} for the definitions of the variables].
		}
		\label{fig_self_sim_sol}
\end{figure}

\begin{figure}[h!]
				\includegraphics[width=.8\linewidth]{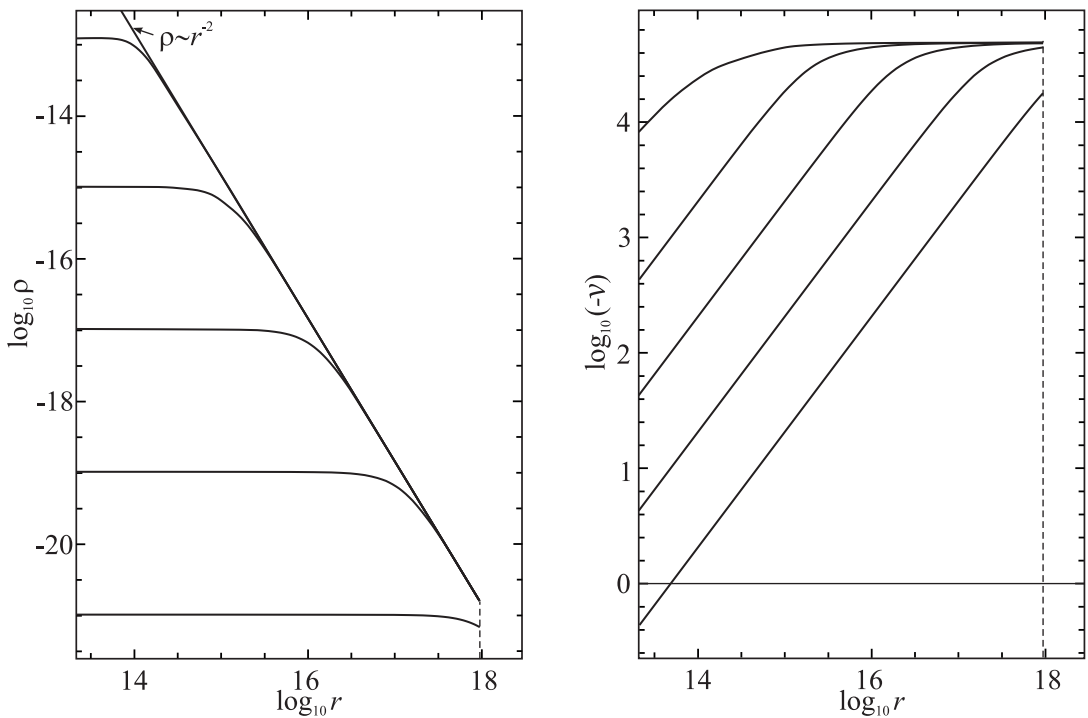}
		\caption{
		Collapse solutions for a one-solar mass configuration with the initial radius $R_i=1\, \text{light-year}=9.46\times 10^{17}\text{cm}$
(shown by the vertical dashed lines).
The density, $\rho$, and the velocity, $-v$,  profiles at
  $t=t_i, 0.1\cdot t_i, 10^{-2}\cdot  t_i, 10^{-3}\cdot  t_i$, and $10^{-4}\cdot t_i$ with the initial $t_i=-3.25 \times 10^{13}\, \text{s}$  shown
   (from bottom to top in both panels).   The dimensions of $r, \rho$, and $v$ are  $\text{cm}$, $\text{g}\, \text{cm}^{-3}$, and $\text{cm s}^{-1}$, respectively.
   As the collapse proceeds,  the density distribution closely approaches the form $\rho\sim r^{-2}$ and the velocity approaches a constant
   [cf. the asymptotic behavior given by Eq.~\eqref{asymp_sol}].
		}
		\label{fig_self_sim_sol_dim}
\end{figure}

Bearing in mind that we would like to have a nonuniform distribution of matter at the end of collapse phase, we plot in
Fig.~\ref{fig_self_sim_sol}  the solutions with a maximally allowed value of
$\alpha_c=0.927$, for which regular solutions still exist. As in the case of the well-known Larson-Penston solutions~\cite{Larson:1969mx,Penston:1969yy}  which are used
in describing the collapse of a system consisting of an isothermal fluid modeled by the polytropic EOS~\eqref{EoS_approx} with $\gamma=1$,
in our case, the collapse starts with an approximately uniform distribution of matter (a fluid cloud) when the pressure gradient is small and the whole cloud begins to collapse practically in free fall.
In the course of time, the density near the boundary of the cloud decreases, while the density in the interior of the cloud increases.
Correspondingly, a pressure gradient appears in the outer part of the cloud, whose presence causes the collapse in this region to be significantly different from a free fall.
For this reason, the density of the fluid rises more rapidly in the interior regions than in the outer parts, and the density distribution becomes peaked at the center.
This is illustrated in Fig.~\ref{fig_self_sim_sol_dim}, which shows that as the collapse proceeds the considerable changes in density occur in the interior regions of the system,
while practically nothing happens in the outer parts of the cloud.

Physically it might be supposed that such a process proceeds up to the formation of some resulting
distribution of matter with a mean density comparable  in order of magnitude with the characteristic density of the static system of Sec.~\ref{stat_conf}.
This, of course,  is a very rough approximation, and one should expect that in reality, somewhere in the process of collapse, the violation of the self-similar regime will take place, and it will
already be necessary to solve the set of partial differential equations~\eqref{eq_cont_2}-\eqref{eq_eul_3}; this would permit one to get a more realistic description of collapse,
as is done, for example, when considering the collapse of an isothermal sphere in Ref.~\cite{Foster1993}.
Nevertheless, even such a rough description permits one to make some preliminary estimates of parameters of the collapse under consideration.

\subsection{Estimation of collapse parameters}

Let us now estimate the parameters of the collapsing systems under consideration. Assume that the collapse starts at some initial instant $t_i<0$
and ends at some final instant $t_f<0$, i.e., the variable $t$ in Eqs.~\eqref{eq_cont_2}-\eqref{polyt_coeff} covers the range $|t_f|\leqslant |t| \leqslant |t_i|$.
For the self-similar motion under consideration, there are two independent dimensional parameters: the initial density, $\rho_i$, and pressure, $p_i$, of the fluid,
where henceforth the index $i$ denotes that the quantities are taken at $t=t_i$. Using them, it is possible to construct a dimensional parameter not containing the unit of
mass, the initial speed of sound  $c_i=\sqrt{\gamma p_i/\rho_i}$. In this case the role of the parameter $k$
appearing in~\eqref{self_variab} will be played by the square of the speed of sound $c_i$. Then, for  $\gamma=2$, we have $k\equiv c_i^2=2 K_i \rho_i$.
Substituting this $k$ in the expression~\eqref{polyt_coeff}, one can find
\begin{equation}
\label{polyt_coeff_gamma_2}
K(t)=\delta K_i \left(\frac{t}{t_f}\right)^2\equiv K_f \left(\frac{t}{t_f}\right)^2,
\end{equation}
where $\delta=8\pi G \rho_i t_f^2$. We put the EOS parameter for the final configuration, $K_f$, equal to the corresponding value which is typical of the
static configuration of Sec.~\ref{stat_conf}. Taking into account the dimensionless variables~\eqref{dmls_var}, one can get
$$
K_f=\frac{1}{2}\frac{\hbar^3}{\mu^4 c},
$$
whose value is determined only by the mass of the spinor field. Then, taking into account the above expressions for $\delta$ and $K_f$ and the relation~\eqref{polyt_coeff_gamma_2}
between $K_i$ and $K_f$, we have
$$
k\equiv 2 K_i \rho_i=\frac{1}{8\pi}\left(\frac{R_L}{t_f}\right)^2.
$$
This expression is used in changing from the dimensionless variables given by Eq.~\eqref{self_variab} to the dimensional physical quantities that have been used
in constructing the graphs shown in Fig.~\ref{fig_self_sim_sol_dim}.

The expression~\eqref{polyt_coeff_gamma_2} describes the evolution of the parameter $K$ in the process of collapse.
When $t=t_i$, this yields $K(t_i)\equiv K_i=\delta K_i\left(t_i/t_f\right)^2$, whence, taking into account the expression for $\delta$, one can find
\begin{equation}
\label{t_init}
t_i=-\frac{1}{\sqrt{8\pi G \rho_i}}.
\end{equation}
Correspondingly, by giving a density of an approximately uniform initial configuration, $\rho_i$, one can find the instant when the collapse begins.
For this instant, we set the radius of the corresponding system equal to half the Jeans length from~\eqref{Jeans_len},
\begin{equation}
\label{R_init}
R_i\equiv\frac{\lambda_{\text{J}}}{2}=\frac{\sqrt{\pi}}{2}\frac{t_i}{t_f}R_L.
\end{equation}
The mass of such initial configuration is then
\begin{equation}
\label{M_init}
M_i=\frac{4\pi}{3}\rho_i R_i^3=\frac{\pi^{3/2}}{48 G}\frac{t_i}{t_f^3}R_L^3,
\end{equation}
where $\rho_i$ is determined from Eq.~\eqref{t_init}.

For simplicity,
we will begin from the assumption that $M_i$ is equal to the mass of the final static configuration
(i.e., there is no mass losses in the process of collapse).
The latter can be expressed in terms of the mean density
\begin{equation}
\label{rho_mean}
\langle\rho\rangle=\frac{3}{4\pi}\frac{M_{s}}{R_s^3}=7.08\times 10^{16}\bar\varepsilon_c\left(\frac{\mu}{\text{GeV}}\right)^4 \text{g cm}^{-3},
\end{equation}
where the mass, $M_s=(\sqrt{\pi}/2) \bar\varepsilon_c M_L$,  and the radius, $R_s=(\sqrt{\pi}/2) R_L$, of the static configuration are taken from Eq.~\eqref{R_M_approx}.
Then we have
$$
M_f\equiv \frac{4\pi}{3}\langle\rho\rangle R_s^3=M_i.
$$
Substituting here $M_i$ from Eq.~\eqref{M_init} and $t_i$ from Eq.~\eqref{t_init}, one can find the expression
$$
t_f=-\frac{1}{\sqrt{8\pi G}}\left(\rho_i\langle\rho\rangle^2\right)^{-1/6}
$$
which determines the value of the final instant when the collapsing system, starting from the initial density $\rho_i$,
would acquire the required mean density from~\eqref{rho_mean}, which is typical of the static configuration.
Since $\rho_i$ is uniquely determined by $t_i$, by giving some value of $t_i$,
one can find $t_f$ and the total collapse time:
\begin{equation}
\label{T_col}
T_{\text{col}}=t_f-t_i=-t_i\left[1-\left(\frac{\rho_i}{\langle\rho\rangle}\right)^{1/3}\right].
\end{equation}

Let us now estimate allowed values of the mass of the spinor field $\mu$. Suppose that we wish to get a configuration with the total mass
$M_{\text{tot}}=n M_\odot$, where $n$ is a constant coefficient. Let us first find a maximally allowed value of $\mu$ that can still provide this total mass of the system.
For this purpose, we make use of the expression for $M_{\text{tot}}$ from Eq.~\eqref{R_M_approx}. Taking into account that the maximum allowed value of $\bar\varepsilon_c\sim 0.1$,
we can obtain
$$
\mu_{\text{max}}\approx\frac{0.38}{\sqrt{n}} \,\text{GeV}.
$$

In turn, to estimate a minimum value of $\mu$ one may start from the fact that the size of the initial configuration cannot be smaller than that of the final system.
The limiting case when they are equal implies that $\rho_i=\langle\rho\rangle$ and $t_i=t_f$. In this case one can find from Eq.~\eqref{R_init}
$$
\mu_{\text{min}}\approx4.75\times 10^{-7}\sqrt{\frac{\text{ly}}{R_i}}\,\text{GeV},
$$
where $R_i$ is measured in light-years. Thus, for the system under consideration, $\mu_{\text{min}}\lesssim\mu\lesssim\mu_{\text{max}}$.
Notice that  $\mu_{\text{max}}$ is determined only by the total mass of the system
(through the coefficient $n$), and $\mu_{\text{min}}$ only by arbitrarily selected sizes of the initial configuration from which the collapse begins.

The total collapse time \eqref{T_col} is determined by the initial density $\rho_i$ and by the mass  $\mu$ (through the mean density $\langle\rho\rangle$).
For some given $\rho_i$, the collapse of a system with $\mu=\mu_{\text{max}}$ takes a maximum time:
$$
T_{\text{col}}^{\text{max}}=-t_i\left(1-1.89\times 10^{-5}n^{2/3}\rho_i^{1/3}\right).
$$
For $\mu<\mu_{\text{max}}$, the time $T_{\text{col}}<T_{\text{col}}^{\text{max}}$, and  $T_{\text{col}}\to 0$ as $\mu\to \mu_{\text{min}}$.
In particular, for a configuration described by the solutions shown in Fig.~\ref{fig_self_sim_sol_dim},
whose total mass is equal to one solar mass ($n=1$) and the initial radius is $R_i=1\, \text{light-year}=9.46\times 10^{17}\text{cm}$,
the initial density
 $\rho_i=5.65\times 10^{-22}\text{g cm}^{-3}$, the initial speed of sound $c_i=6.55\times 10^{3}\text{cm s}^{-1}$,
 and $T_{\text{col}}^{\text{max}}=3.25\times 10^{13}\,\text{s}\approx 10^6\,\text{years}$.

\section{Conclusion}
\label{conclus}

In the present paper, we have been continuing our investigations of the Dirac stars begun in Refs.~\cite{Dzhunushaliev:2018jhj,Dzhunushaliev:2019kiy,Dzhunushaliev:2019uft}.
In doing so, we have studied  nonrelativistic gravitating configurations consisting of a spinor fluid modeled by the effective equation of state~\eqref{EoS_eff}
which approximately describes the limiting configurations supported by
two nonlinear spinor fields~\cite{Dzhunushaliev:2018jhj}. For this case, we have found a two-parametric family of static regular solutions
describing compact systems whose sizes and total masses can vary within a wide range depending on
the mass of the spinor field and on the central density of the fluid.

It is shown that, in Newtonian gravitational theory, the EOS~\eqref{EoS_eff} may be replaced to good accuracy by an approximate expression corresponding
to a polytropic fluid with $p\sim \rho^2$. This allowed us to derive the corresponding self-similar equations describing a motion of the spinor fluid in its own gravitational field.
These equations have been used to model the collapse process of an initial, approximately uniform, cloud consisting of the spinor fluid.
As a result of the collapse, there arises some nonuniform distribution of matter with $\rho\sim r^{-2}$
that may serve as a core whose subsequent non-self-similar evolution may eventually lead to
the creation of a static equilibrium object  of the type considered in Sec.~\ref{stat_conf}. We have estimated the general characteristics of the collapse and also,
by giving a particular mass of the initial cloud and its density, presented the corresponding results of the calculations.

\section*{Acknowledgments}
The work was supported by the Program No.~BR10965191 of the Ministry of Education and Science of the Republic of Kazakhstan.
We are also grateful to the Research Group Linkage Programme of the Alexander von Humboldt Foundation for the support of this research.


\begin{thebibliography}{999}
\bibitem{Schunck:2003kk}
  F.~E.~Schunck and E.~W.~Mielke,
  General relativistic boson stars,
  Classical\ Quantum\ Gravity\  {\bf 20}, R301 (2003).

\bibitem{Liebling:2012fv}
  S.~L.~Liebling and C.~Palenzuela,
  Dynamical Boson Stars,
  Living Rev.\ Relativity\  {\bf 15}, 6 (2012);
   {\bf 20}, 5 (2017).

\bibitem{Brito:2015pxa}
  R.~Brito, V.~Cardoso, C.~A.~R.~Herdeiro, and E.~Radu,
  Proca stars: Gravitating Bose–Einstein condensates of massive spin 1 particles,
  Phys.\ Lett.\ B {\bf 752}, 291 (2016).

\bibitem{Herdeiro:2017fhv}
  C.~A.~R.~Herdeiro, A.~M.~Pombo, and E.~Radu,
  Asymptotically flat scalar, Dirac and Proca stars: discrete vs. continuous families of solutions,
  Phys.\ Lett.\ B {\bf 773}, 654 (2017).

\bibitem{Sanchis-Gual:2019ljs}
  N.~Sanchis-Gual, F.~Di Giovanni, M.~Zilhão, C.~Herdeiro, P.~Cerdá-Durán, J.~A.~Font, and E.~Radu,
  Non-linear dynamics of spinning bosonic stars: formation and stability,
  Phys.\ Rev.\ Lett.\  {\bf 123}, 221101 (2019).

  \bibitem{Bartnik:1988am}
  R.~Bartnik and J.~Mckinnon,
  Particle - Like Solutions of the Einstein Yang-Mills Equations,
  Phys.\ Rev.\ Lett.\  {\bf 61}, 141 (1988).

\bibitem{Volkov:1998cc}
  M.~S.~Volkov and D.~V.~Gal'tsov,
  Gravitating nonAbelian solitons and black holes with Yang-Mills fields,
  Phys.\ Rep.\  {\bf 319}, 1 (1999).

\bibitem{Finster:1998ws}
  F.~Finster, J.~Smoller, and S.~T.~Yau,
  Particle - like solutions of the Einstein-Dirac equations,
  Phys.\ Rev.\ D {\bf 59}, 104020 (1999).

\bibitem{Finster:1998ux}
  F.~Finster, J.~Smoller, and S.~T.~Yau,
  Particle - like solutions of the Einstein-Dirac-Maxwell equations,
  Phys.\ Lett.\ A {\bf 259}, 431 (1999).

\bibitem{Krechet:2014nda}
  V.~G.~Krechet and I.~V.~Sinilshchikova,
  Self-Gravitating Nonlinear Spinor Field in Stationary Spaces with Spherical Symmetry,
  Russ.\ Phys.\ J.\  {\bf 57}, 870 (2014).

 \bibitem{Adanhounme:2012cm}
  V.~Adanhounme, A.~Adomou, F.~P.~Codo, and M.~N.~Hounkonnou,
  Nonlinear spinor field equations in gravitational theory: spherical symmetric soliton-like solutions,
  J.\ Mod.\ Phys.\  {\bf 3}, 935 (2012).

\bibitem{Bronnikov:2019nqa}
  K.~A.~Bronnikov, Y.~P.~Rybakov, and B.~Saha,
  Spinor fields in spherical symmetry: Einstein–Dirac and other space-times,
  Eur.\ Phys.\ J.\ Plus {\bf 135}, no. 1, 124 (2020).

\bibitem{Bronnikov:2004uu}
  K.~A.~Bronnikov, E.~N.~Chudaeva, and G.~N.~Shikin,
  Self-gravitating string-like configurations of nonlinear spinor fields,
  Gen.\ Relativ.\ Gravit.\  {\bf 36}, 1537 (2004).

\bibitem{Bronnikov:2009na}
  K.~A.~Bronnikov and J.~P.~S.~Lemos,
  Cylindrical wormholes,
  Phys.\ Rev.\ D {\bf 79}, 104019 (2009).

\bibitem{Ribas:2010zj}
M.~O.~Ribas, F.~P.~Devecchi, and G.~M.~Kremer,
Fermionic cosmologies with Yukawa type interactions,
Europhys.\ Lett.\  {\bf 93}, 19002 (2011).

\bibitem{Ribas:2016ulz}
M.~O.~Ribas, F.~P.~Devecchi, and G.~M.~Kremer,
Cosmological model with fermion and tachyon fields interacting via Yukawa-type potential,
  Mod.\ Phys.\ Lett.\ A {\bf 31}, 1650039 (2016).

\bibitem{Saha:2016cbu}
B.~Saha,
Nonlinear spinor field in isotropic space-time and dark energy models,
Eur.\ Phys.\ J.\ Plus {\bf 131}, 242 (2016).

\bibitem{Dzhunushaliev:2018jhj}
  V.~Dzhunushaliev and V.~Folomeev,
Dirac stars supported by nonlinear spinor fields,
  Phys.\ Rev.\ D {\bf 99}, 084030 (2019).

\bibitem{Dzhunushaliev:2019kiy}
  V.~Dzhunushaliev and V.~Folomeev,
Dirac star in the presence of Maxwell and Proca fields,
  Phys.\ Rev.\ D {\bf 99}, 104066 (2019).

\bibitem{Dzhunushaliev:2019uft}
  V.~Dzhunushaliev and V.~Folomeev,
Dirac Star with SU(2) Yang-Mills and Proca Fields,
  Phys.\ Rev.\ D {\bf 101}, no. 2, 024023 (2020).

\bibitem{Prialnik}
D.~Prialnik, {\it An Introduction to the Theory of Stellar Structure and Evolution} (Cambridge University Press, Cambridge, 2010).

\bibitem{Choptuik:1992jv}
  M.~W.~Choptuik,
Universality and scaling in gravitational collapse of a massless scalar field,
  Phys.\ Rev.\ Lett.\  {\bf 70}, 9 (1993).

\bibitem{Gonzalez:2008xk}
  J.~A.~Gonzalez, F.~S.~Guzman and O.~Sarbach,
Instability of wormholes supported by a ghost scalar field. II. Nonlinear evolution,
  Class.\ Quant.\ Grav.\  {\bf 26}, 015011 (2009).

\bibitem{Ventrella:2003fu}
  J.~F.~Ventrella and M.~W.~Choptuik,
Critical phenomena in the Einstein massless Dirac system,
  Phys.\ Rev.\ D {\bf 68}, 044020 (2003).

\bibitem{Folomeev:2011uj}
  V.~Folomeev,
Nonrelativistic isothermal fluid in the presence of a chameleon scalar field: static and collapsing configurations,
  Phys.\ Rev.\ D {\bf 85}, 024008 (2012).


\bibitem{Zeld}
Ya. B. Zel'dovich and  I. D. Novikov, {\it Stars and Relativity} (Dover, New York, 1996).

\bibitem{Narain:2006kx}
  G.~Narain, J.~Schaffner-Bielich and I.~N.~Mishustin,
Compact stars made of fermionic dark matter,
  Phys.\ Rev.\ D {\bf 74}, 063003 (2006).

\bibitem{Suto:1988}
Y.~Suto and J.~Silk, Self-similar dynamics of polytropic gaseous spheres,
  Astrophys.\ J.\  {\bf 326}, 527 (1988).

\bibitem{Shu:1977uc}
  F.~H.~Shu,  Self-similar collapse of isothermal spheres and star formation,
  Astrophys.\ J.\  {\bf 214}, 488 (1977).

\bibitem{Hunter:1977}
C.~Hunter,  The collapse of unstable isothermal spheres,
  Astrophys.\ J.\  {\bf 218}, 834 (1977).

\bibitem{Larson:1969mx}
  R.~B.~Larson,
  Numerical calculations of the dynamics of collapsing proto-star,
  Mon.\ Not.\ Roy.\ Astron.\ Soc.\  {\bf 145}, 271 (1969).

\bibitem{Penston:1969yy}
  M.~V.~Penston,
  Dynamics of self-gravitating gaseous spheres-III. Analytical results in the free-fall of isothermal cases,
  Mon.\ Not.\ Roy.\ Astron.\ Soc.\  {\bf 144}, 425 (1969).

\bibitem{Foster1993}
P.~Foster and R.~Chevalier, Gravitational collapse of an isothermal sphere,
  Astrophys.\ J.\  {\bf 416}, 303 (1993).

\end{thebibliography}
\end{document}